\documentstyle[multicol,aps,prl]{revtex}

\begin{document}
\widetext
\draft

\title{Spectra and statistics of velocity and temperature fluctuations 
                  in turbulent convection.}
                      
\author{ Shay Ashkenazi and Victor Steinberg}

\address{Department of Physics of Complex Systems\\
         The Weizmann Institute of Science, 76100 Rehovot, Israel}

\date{\today}
\maketitle
\begin{abstract}
Direct measurements of the velocity in turbulent convection of $SF_6$ near 
its 
gas-liquid critical point by light scattering on the critical density 
fluctuations  were conducted. The temperature, velocity and cross frequency power
 spectra in a 
wide range of the Rayleigh and Prandtl numbers show scaling behaviour
with indices, which are rather close to the 
Bolgiano-Obukhov scaling in the wavenumber domain. The statistics 
of the velocity fluctuations remains Gaussian up to the Reynolds numbers 
$10^5$.\\

\end{abstract}

\begin{multicols}{2}

\narrowtext
Turbulent convection exemplifies many startling aspects of turbulent flows.
The discovery of scaling laws in the heat transport and temperature
 statistics shed new light on the nature of the convective turbulence
 \cite{hesl,wu}. Bursts of thermal plumes from 
 thermal boundary layers and a coherent large scale circulation, which 
 modifies the boundary layers via its shear, are found to coexist in a 
 closed convection cell\cite{hesl,wu,cilib}. These two salient features which
  reflect 
 the competition between buoyancy and shear, were used in theoretical models
 to explain the observed scalings in the temperature field\cite{hesl,sig,sig1}. 
 The predictions 
 which follow from these models for the velocity field, are still far from an
  experimental verification\cite{sig1}.\\ 
  Therefore direct measurements of the velocity field would become a crucial
   step
to test predictions of various theoretical models, to understand the mechanism 
of energy vs. entropy transfer, to determine relative contribution of the 
thermal plumes and the large-scale circulation to the heat transport, and 
to study the mechanism of an onset and generation of the
large-scale flow by intermittent thermal bursts.\\
  Lack of the velocity field measurements is explained by the fact that both 
 conventional methods such as a hot-wire anemometry and a laser Doppler 
 velocimetry(LDV), are not suitable 
 for turbulent convection, as was commonly agreed\cite{tong}.
 Indeed a hot-wire anemometry is useless due to strong temperature 
fluctuations which accompany velocity fluctuations, and thus do not permit to
 correctly measure velocity. Fluctuations in a refractive index due to the
 large temperature fluctuations  corrupt a signal and to drastically reduce a
signal-to-noise ratio of LDV measurements because of  wandering and 
defocusing of  laser 
 beams. Various attempts to measure the velocity field were rather
 limited due to severe experimental difficulties to directly measure the fluid 
velocity in convective flows in a wide dynamical range \cite{tong,til}.  
Novel direct velocity measurements by
dual- and single-beam two-color cross-correlation spectroscopy technique
 \cite{tong} in rather limited range of the Rayleigh numbers, $Ra$, still 
 do not provide information about the velocity spectra 
even in the frequency domain and direct information about statistics in a 
sufficiently wide dynamical range.\\   
Another problem, which is still controversial and requires further experimental 
test, is whether the scaling laws in the inertial range of the temperature and 
velocity power spectra in homogeneous turbulent convection are either those 
suggested 
by Kolmogorov, or those suggested by Bolgiano and Obukhov(BO)\cite{sig,proc}. 
It was 
argued\cite{sig}, that only the Kolmogorov scaling $k^{-5/3}$ for both the 
temperature and velocity power spectra, where $k$ is the wavenumber, is 
consistent. It means that the temperature is a passive scalar which just 
follows the
 velocity field. Other theories predict that depending on the parameters 
 range 
 either the Kolmogorov or BO scalings can be observed
 \cite{proc}. The latter gives a power law $k^{-11/5}$ for the velocity 
 fluctuations spectrum, and $k^{-7/5}$ for the temperature fluctuations 
 spectrum.
Here the temperature is not a passive scalar but provides a potential energy 
on all scales. First measurements of the frequency spectra of the temperature 
fluctuations\cite{hesl} and then many subsequent experiments\cite{cilib,sig1} reveal
 the BO 
power law, although the inertial range was very limited, and the applicability
 of the Taylor hypothesis is not completely justified. The only measurements of
 the velocity spectra in a real space by a homodyne spectroscopy method in 
 rather narrow range of $Ra$ also confirm the BO predictions\cite{shen}. 
 However, various
 numerical simulations and, particularly recent remarkable ones \cite{borue} 
 support the Kolmogorov scaling and call for further experiments.\\
Our experimental studies during the last several years convincingly
 demonstrated that the Rayleigh-Benard convection in a gas near the gas-liquid
 critical point (CP) is 
a very appropriate system to study high Ra turbulent convection
\cite{shay,pap}. Singular
 behaviour of the thermodynamic and kinetic properties
of the fluid near $T_c$ provides the opportunity to reach both extremal
 values of the
control parameter (Ra up to $5\cdot 10^{14}$ has been reached in our 
experiment on  ${SF}_6$  )  and to scan Prandtl number,
$Pr$, over an extremely wide range ( from one till several hundred). All these 
features make the system unique in  this respect\cite{shay,pap}.   
 However, the most relevant aspect of the system to the results presented 
 in this Letter, is our finding that the presence of the critical 
fluctuations provides us the possibility to conduct LDV
 measurements in a rather wide range of proximity to CP between $3\cdot 10^{-4}$
  and $10^{-2}$ in the reduced temperature 
  $\tau=(\overline T-T_c)/T_c$, where $\overline T$ is the mean cell
   temperature. It 
 corresponds to variation in $Pr$ from about 20 till 200.
 The upper limit in $\tau$ is defined by a  
 scattered light intensity. It is experimentally found that at $\tau > 10^{-2}$
  the photomultiplier current is too low to recover the 
  Doppler frequency shift.
  The lower limit is determined by multiple scattering in a large 
  convection cell as a result of which the light intensity at the detector 
  is drastically reduced. The signal-to-noise ratio of the LDV measurements is 
  increased with the closeness to CP until it drops rather abruptly at 
  $\tau < 3\cdot 10^{-4}$\cite{shay,pap}.\\ 
 Critical density fluctuations moving with the flow, scatter light as 
 effectively as seeding particles in a conventional LDV: the corresponding 
 signal from a photomultiplier is of excellent quality\cite{shay}.
 The light scattering occurs
  on the short-scale thermal density fluctuations advected by the large scale
  turbulent fluctuations. In the range of
$\tau$ used in the experiment, the scattering takes place mostly on the thermal 
   density fluctuations. Indeed, due to smallness of the Mach number at 
    maximum flow velocities  the
    density fluctuations due to turbulent flow were negligible.\\
     It is well known
   from conventional LDV measurements on seeding particles,
 that the presence of too many seeding particles in a scattering 
 volume leads to significant corruption of the signal. So, the question is, 
 how  so large density of scatterers provides so clean signal. The light 
 scattering on the critical fluctuations occurs differently from that on
 the seeding  particles: the scattering amplitude is random in both the 
 amplitude absolute value and phase in space and time. The crucial point here
  is that the scattering occurs on sound phonons due to large fluid compressibility,
  rather than on individual fluctuations of the order of the correlation length.
  So the time variations of the pressure fluctuations are correlated even at great distances.
  In the range of $\tau$ mentioned above, the correlation 
 length is less than $10^{-5}$ cm. On the other hand, the wavenumber of 
 the light excited sound phonons is $q=2k_0\sin\theta/2$, where $\theta$ is the 
 angle between the crosssed laser beams in the LDV system, and 
 $k_0$
 is the light wavenumber. Then the corresponding sound wavelength
  is  $\lambda=2\pi/q\approx 3 \mu m$, and the frequency 
  $f=qc_s \approx 20$ Mhz, where $c_s$ is the sound velocity. At this frequency and 
  in this range of $\tau$ 
  the sound attenuation length will be by orders of magnitude larger than the 
  correlation length\cite{atten}, and turns out to be the length on which the coherent
   light scattering takes place. Since it is also less or of the order 
   of the beam waist in the scattering volume, it means that at these
    conditions the scattering  occurs 
    just on a few coherent scattering regions like
    on a few particles. That is the reason for observation of a good quality
    signal. Of course, the LDV used the scatterred signal rather close to the central
    Rayleigh line, at frequency differences much smaller than the corresponding
    Mandelshtam-Brillouin doublet frequency shift  
     due to the 
    phonon scattering $\Delta \omega=f$ (several kHz compared with 20 MHz)\cite{land}.
     But nonlinear 
    coupling between the amplitude and the phase of the scattering light produces
    sufficient signal at low frequency shifts to be observable.\\
  In this Letter we studied $Pr$ dependence of spectra and  statistical
  properties 
 of the vertical velocity fluctuations at one location in the frequency 
 domain using this
 novel LDV technique. In addition we also obtained the temperature fluctuations
 spectra in
  a wide range of $Ra$ and $Pr$.\\
 The experiment we present here, was done with a high purity gas 
$SF_6$ (99.998\%) in the vicinity of $T_c$ and at the 
critical density($\rho_c=730 kg/m^3$). This fluid was chosen due to the 
relatively low critical temperature
 ($T_c=318.73$ K) and  pressure ($P_c=37.7$ bar) and well-known 
thermodynamic and kinetic properties far away and in the vicinity of 
CP. The set-up was described elsewhere\cite{shay,pap}. The cell is a box of
 a cross-section
76x76 $mm^2$ formed by 4 mm plexiglass walls, which are sandwiched 
between a Ni-plated 
mirror-polished copper bottom plate and a 19 mm thick sapphire top plate of 
$L=105$ mm apart. \\ 
The cell is placed inside the pressure vessel with two side thick plastic 
windows to withstand the pressure difference up to 100 bar. So the cell had 
 optical accesses from above through the sapphire window and from the sides. 
 They were used for both shadowgraph flow visualization  
  and for LDV. The pressure vessel was placed
inside a water bath which  was stabilized with rms temperature fluctuations at 
 the level of 0.4 $mK$. The gas pressure was continuously measured 
 with 1 $mbar$ resolution by the absolute pressure gauge. Together with
calibrated 100$\Omega$ platinum resistor thermometer they provide us the
thermodynamic scale to define the critical parameters of the fluid and then to
use the parametric equation of state developed recently for $SF_6$
\cite{seng}.\\
  Local temperature measurements in a gas were made by three $125\mu m$ 
  thermistors suspended on glass fibers in the  interior of the cell 
  (one at the center, and two about half way from the wall). Local vertical 
  component velocity measurements at about $L/4$ from the bottom plate were 
 conducted by using LDV on the critical density fluctuations\cite{shay,pap}.\\ 
  A typical time series of the vertical velocity component  of turbulent 
  convection at $Pr=93$ and the  power spectra at $Ra$ from $3\cdot 10^{11}$ 
  to $8\cdot 10^{13}$ are 
 shown in Fig.1. The spectra for other values of $Pr$ look similar. The 
 spectra 
 at higher $Ra$ are characterized by well-defined low frequency 
 peaks related to the large scale circulation, by surprisingly large inertial
range, and by relatively high level of white noise which significantly 
limits
the dynamical range of the spectra. The noise level does not depend on the flow 
 velocity and remains at the same level regardless of the heating power which 
 drives the convection flow, but does depend on $\tau$. So the  
 signal-to-noise ratio increases with approaching the CP
  and with a
 velocity increase. As was shown, the source of the white noise is the phase 
 noise due to the critical density fluctuations. The results of these
 studies will be published elsewhere\cite{shay}.\\
The random phase of the scattering signal causes an additional noise in the
 velocity measurements, and thus limits the obtaining the reliable data. Indeed, 
 the measured Doppler frequency shift can be presented as 
 $\dot{\phi}=\vec{V}\cdot\vec{k}+\dot{\phi_{th}}$, where $\vec{V}\cdot\vec{k}$
 is the velocity component along the wavevector $\vec{k}$, and $\dot{\phi_{th}}$
 is the frequency shift due to the thermal phase noise. Then the corelation function 
 of the measured frequency shift, averaged over long time, is 
 $\langle \dot{\phi}\dot{\phi}\rangle=\langle (\vec{V}\cdot\vec{k})^2 \rangle +
 \langle \dot{\phi_{th}}\dot{\phi_{th}}\rangle +2\langle (\vec{V}\cdot\vec{k}
 \cdot\dot{\phi_{th}}\rangle$.
 This expression at adiabatically slow temporal variations of 
 $\vec{V}\cdot\vec{k}$ reduces to 
 $\langle \dot{\phi}\dot{\phi}\rangle=\langle (\vec{V}\cdot\vec{k})^2 \rangle+
 \langle \dot{\phi_{th}}\dot{\phi_{th}}\rangle$, that leads to the same additive 
 relation in the corresponding frequency spectrum. So, at 
 $\vec{V}\cdot\vec{k}\gg\dot{\phi_{th}}$ one reproduces the velocity from the
  measured Doppler frequency shift.  Besides, the 
  smallest energy containing eddies should be larger than both the correlation 
  length and beam waist. If these conditions are satisfied, one can expect
  that the frequency shift of the scattered light is equal to the fluid 
  velocity in the wavenumber units.\\
 From the fit in  the inertial range (Fig.1) one finds well-defined power law 
 behaviour. In fact the LDV measurements do not include the whole inertial 
range  due to the noise limitation. So one cannot reach the dissipation region.
 The scaling index obtained by the linear fit, similar to that shown in Fig.1, and
  averaged over all values of $Ra$ and $Pr$, gives $-2.4\pm .2$, which rather 
 close to the BO scaling  in the wavenumber domain\cite{proc}.\\
 From the velocity time series the probability density function (PDF) of the
  velocity fluctuations
normalized by the rms velocity, was also constructed. The PDF of the 
velocity 
 fluctuations as well as the velocity differences at a finite time lag were 
 Gaussian (Fig.2) for all $Ra$ and $Pr$ under studies. We investigated also  
 the dependence of the rms vertical velocity fluctuations and mean vertical
  velocity on $Ra$ and $Pr$. The latter scaling behaviour was reported 
  elsewhere\cite{shay,pap}. The former one is presented in Fig.3 in 
$Re(=V_sL/\nu)$ vs $Ra$ coordinates. Here $V_s$ is the rms of vertical 
velocity 
  fluctuations, and $\nu$ is the kinematic viscosity. The fit to the data for 
 three values of $Pr=27,45$, and $93$ suggest the scaling 
 $Re=0.005 Ra^{0.43 \pm .02}$ 
  with no $Pr$ dependence. The data for $Pr=190$ deviate from this scaling 
 behaviour and can be better fitted with the scaling exponent 0.34. The former 
  scaling exponent of $Re$ vs $Ra$ for lower $Pr$ is rather close to the value
 obtained for water($Pr=7$) and for the theoretical prediction 3/7 of both the 
 mixing zone\cite{hesl} and the large scale flow theories\cite{sig}. We would 
 like to emphasize that $Pr$ dependence of $Re$ in the bulk is very different 
 from that found for the large scale circulation\cite{shay,pap}.\\
 The power spectra of the temperature fluctuations were measured both at the 
 cell center and on the side( $L/4$ from the wall)
 in a wide range of $Ra$ and $Pr$. We used also the data which we
took in the compressed gas $SF_6$ far away from CP at $P=20 bar$, 
$\overline T=303 K$ and the density 
 $\rho=0.18 g/cm^3$, that corresponds to $Pr=0.9$, and at $P=50 bar$,
  $\overline T=323 K$ and $\rho=1.07 g/cm^3$, that corresponds to $Pr=1.5$. 
  The data far away from CP cover the range of $Ra$ between $10^9$ and 
 $5\cdot 10^{12}$. Most of the power spectra did not show any reasonable range 
  of a scaling law. At $Pr$ above 50 and the highest $Ra$ the power law 
  behaviour became noticable. At $Pr>90$ and the highest values of $Ra$ 
  the scaling region of about one decade in frequency was observed (Fig.4).
  The scaling index was obtained by fitting the spectra by\cite{wu} 
  
  $$P(f)=\left\{ \begin {array}{ll}
             Bf^s & \mbox{if $f<f_c$}\\
          Bf^s(f/f_c)^{ln(f/f_c)^{-\alpha}} & \mbox{if $f>f_c$}\end{array}\right\}$$

 where $s,f_c$, and $\alpha$ are the fit parameters. The scaling index
  obtained 
 by the fit and averaged over all spectra, is $s=-1.45 \pm 0.1$, that is close
 to the BO power law for the temperature fluctuations spectra in the wavenumber
 presentation.\\            
  Thus the measured scaling indices for both the velocity and temperature
   fluctuations in the frequency domain are rather close to the theoretically
    predicted BO scaling
 in the wavenumber presentation. As commonly accepted, the Taylor 
 hypothesis can be used to pass from the wavenumber to frequency domain. 
 Since we measured both the velocity and temperature spectra in the reqion 
 with non-zero mean velocity, the application of the Taylor hypothesis is
  better justified.
  From the velocity and temperature time seria taken at the same spatial location,
   we constructed cross-correlator which in the Fourier frequency domain 
   also showed scaling behaviour with the 
 scaling index $-1.85 \pm 0.1$, that is
  close to the theoretically predicted value in the BO region\cite{proc}.\\

  This work was partially supported by the Minerva Foundation and the Minerva 
 Center for Nonlinear Physics of Complex Systems. We are thankful to V. Lebedev
 for illuminating and fruitful 
   discussions, criticism and useful suggestions. VS is greatful 
   for support of the Alexander von Humboldt Foundation.\\

\begin{figure}
\caption{Power spectra of velocity fluctuations for $Ra$ from $3\cdot 10^{11}$ 
to $8\cdot 10^{13}$ at $Pr=93$. Solid line is a power law fit. The insert: 
Vertical velocity time series at $Ra=3\cdot 10^{12}$. }

\label{figa}
\end{figure}

\begin{figure}
\caption { PDF of the velocity time series for $Ra$ from $3\cdot 10^{11}$ 
to $8\cdot 10^{13}$ at $Pr=93$. Each PDF is normalized by the rms of each time 
series. The thick line is a Gaussian PDF.}

\label{figb}
\end{figure}

\begin{figure}
\caption { $Re$ for the rms vertical velocity fluctuations as a function
  of $Ra$ for three values of $Pr$: circles-27, up-triangles-45,
  down-triangles-93. Solid line is a power law fit.}

\label{figc}
\end{figure}

\begin{figure}
\caption {Power spectra of temperature fluctuations at the cell center at
$Pr=93$, $Ra=8.4\cdot 10^{13}$-top; $Pr=190$, $Ra=1.5\cdot 10^{14}$-middle;
 $Pr=300$, $Ra=3\cdot 10^{14}$-bottom. In each plot a line of the best fit
 by the model described in the text, is shown.}

\label{figd}
\end{figure}

\end{multicols}

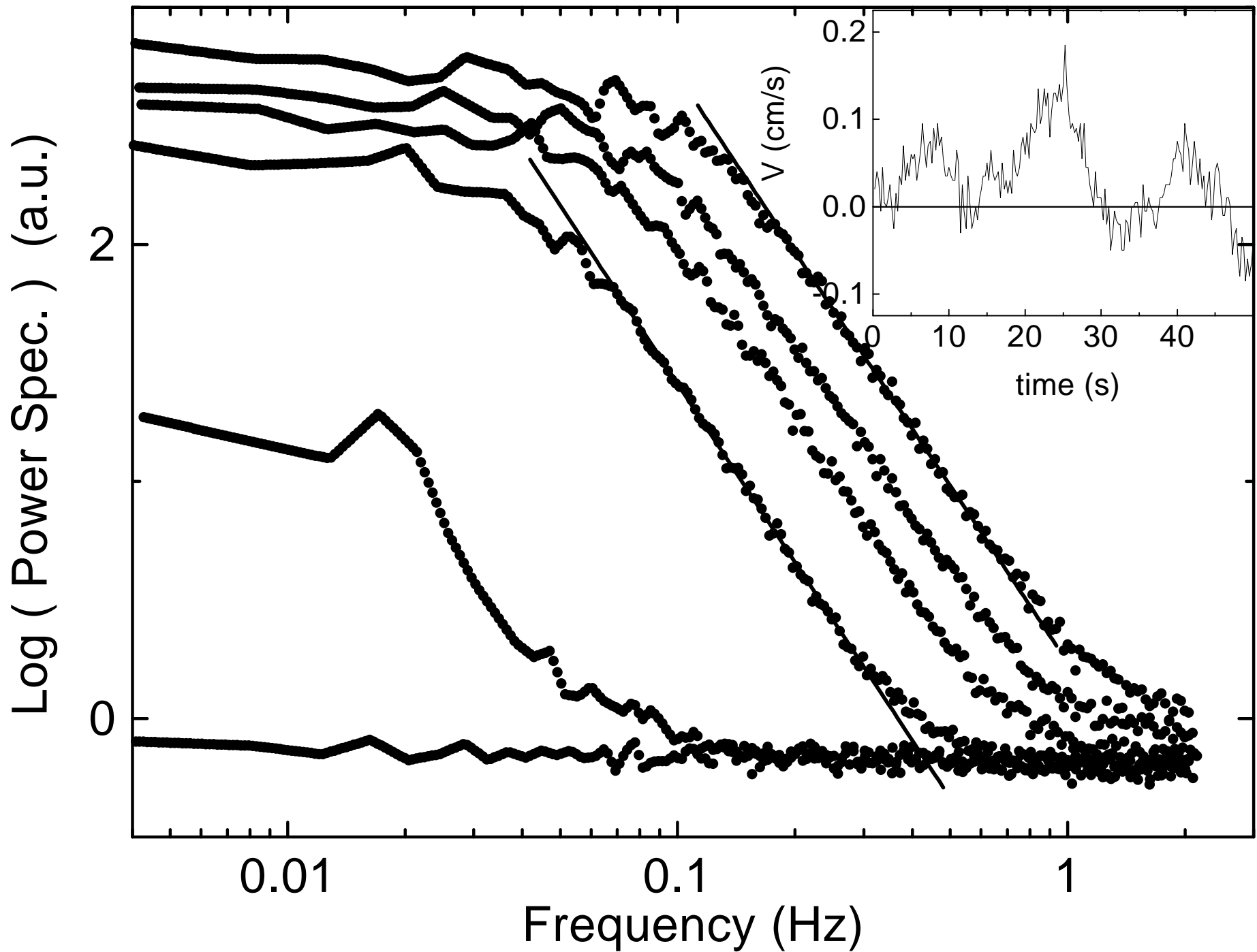
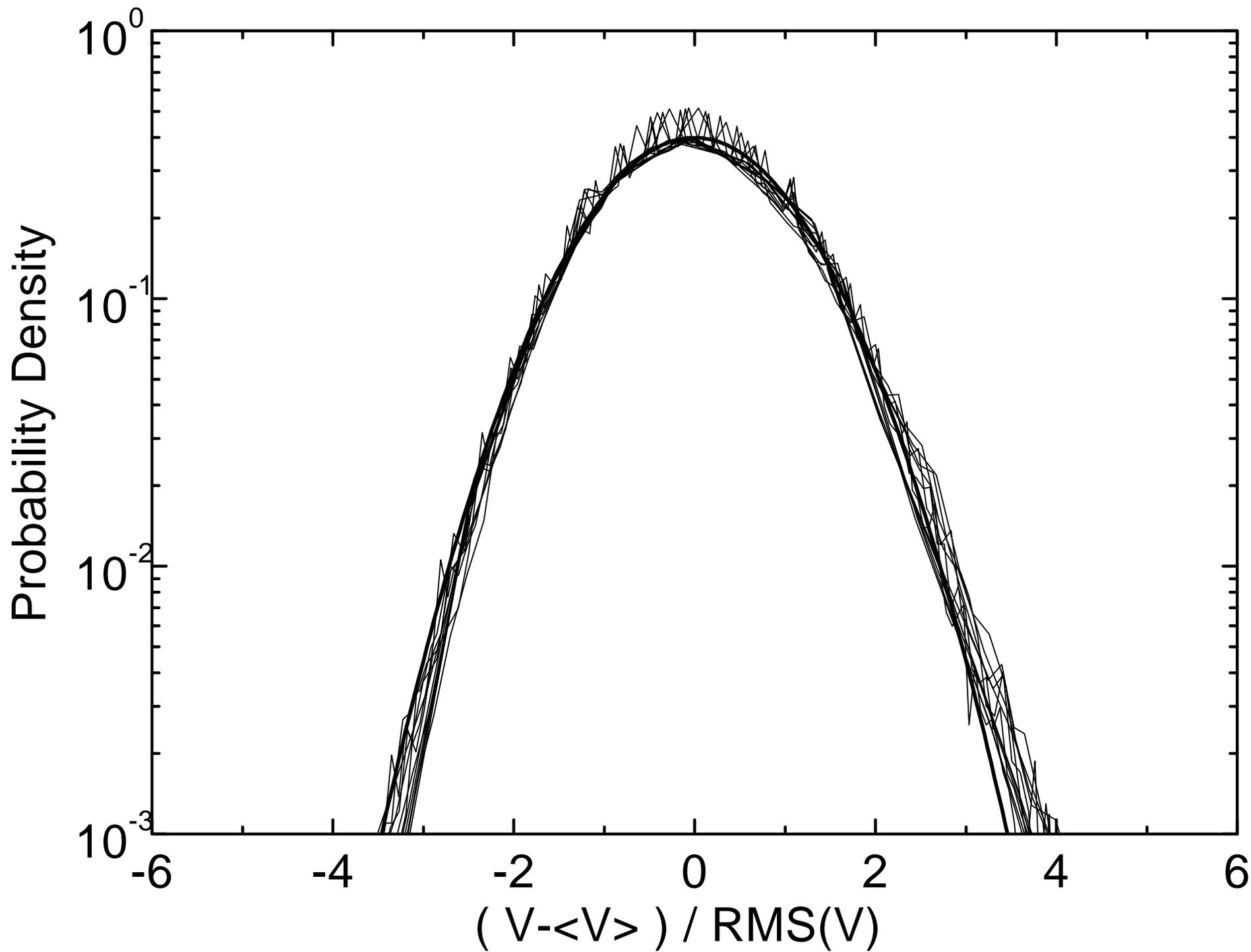
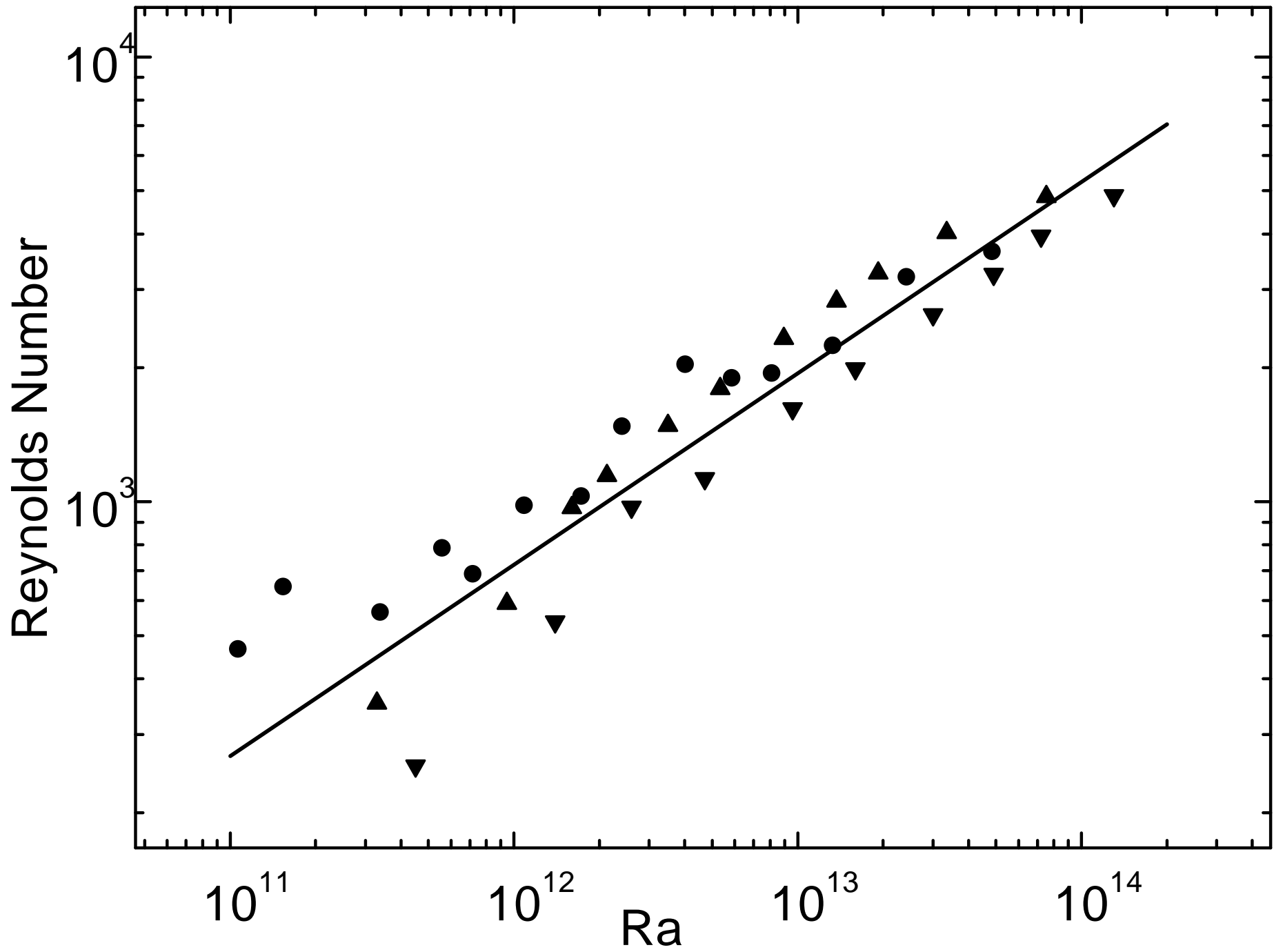
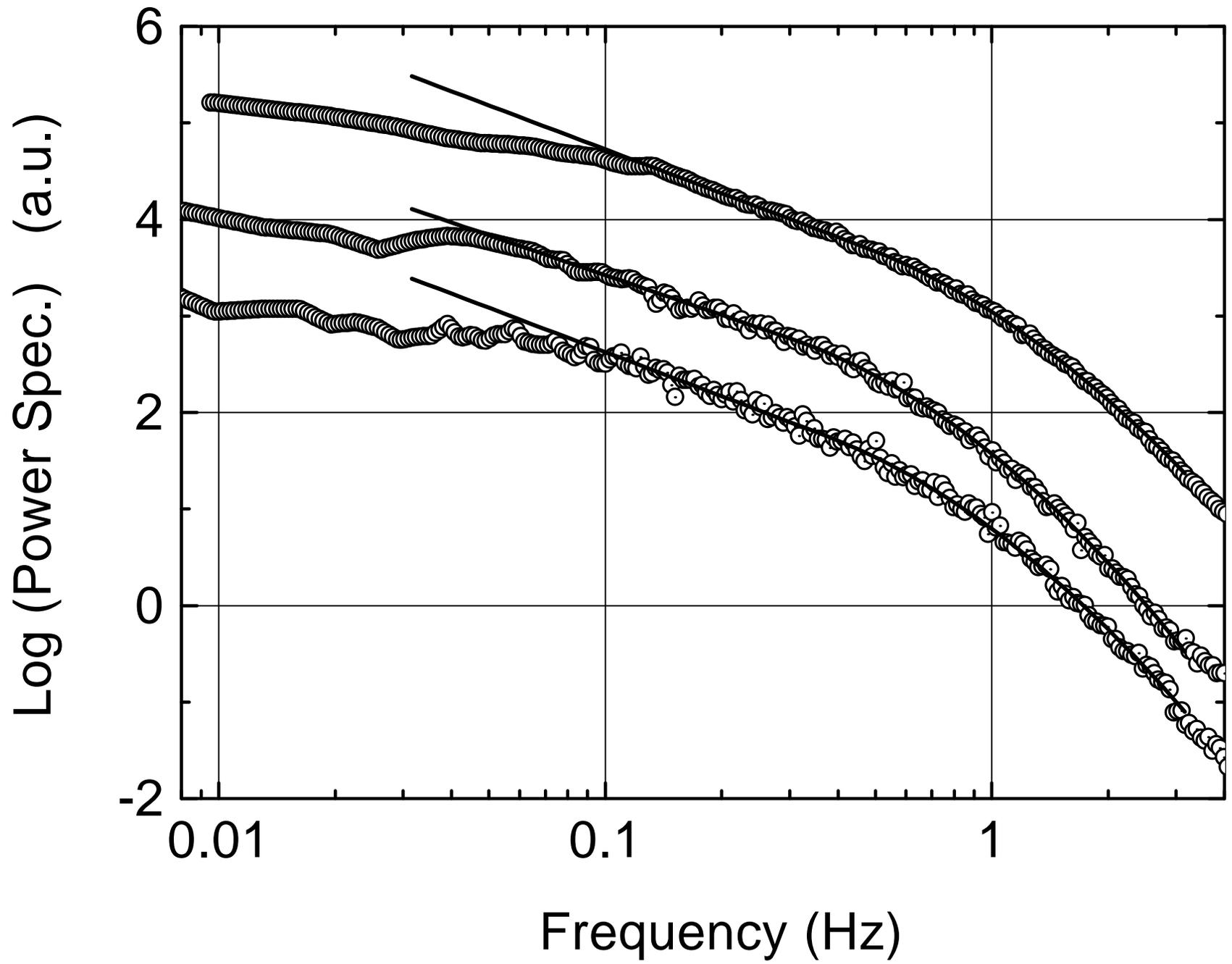
\end{document}